\documentclass[twocolumn,aps,prl,superscriptaddress,longbibliography]{revtex4-2}
\usepackage{graphicx}
\usepackage{amsmath}
\usepackage{amssymb}
\usepackage{braket}
\usepackage{physics}
\usepackage{algpseudocode}
\usepackage[colorlinks=true,linkcolor=blue,citecolor=blue,urlcolor=blue]{hyperref}

\begin{document}

\title{AI-Enabled Qubit Loss Detection and Replenishment Strategy for Atom Array Quantum Computing}
\author{Yuqing Wang}
\thanks{These authors contributed equally to this work.}
\affiliation{Department of Physics and State Key Laboratory of Low Dimensional Quantum Physics, Tsinghua University, 100084, Beijing, China.}
\affiliation{Beijing Key Laboratory of Cold Atom Quantum Computation, Tsinghua University, 100084, Beijing, China.}
\author{Xiaotian Nie}
\thanks{These authors contributed equally to this work.}
\affiliation{Intelligent Quantum Inception Co., Ltd., Haidian, Beijing, 100083, China.} 
\affiliation{Beijing Key Laboratory of Cold Atom Quantum Computation, Tsinghua University, 100084, Beijing, China.}
\author{Jiale Dai}
\affiliation{Chen-Ning Yang Institute for Advanced Study, Tsinghua University, Beijing 100084, China.}
\affiliation{Beijing Key Laboratory of Cold Atom Quantum Computation, Tsinghua University, 100084, Beijing, China.}
\author{Zhongyi Ni}
\affiliation{Hong Kong University of Science and Technology (Guangzhou), Guangzhou 511453, China.}
\author{Tao Zhang}
\email{tao\_zhang@mail.tsinghua.edu.cn}
\affiliation{Chen-Ning Yang Institute for Advanced Study, Tsinghua University, Beijing 100084, China.}
\affiliation{Intelligent Quantum Inception Co., Ltd., Haidian, Beijing, 100083, China.}
\affiliation{Beijing Key Laboratory of Cold Atom Quantum Computation, Tsinghua University, 100084, Beijing, China.} 
\author{Hui Zhai}
\email{hzhai@tsinghua.edu.cn}
\affiliation{Chen-Ning Yang Institute for Advanced Study, Tsinghua University, Beijing 100084, China.}
\affiliation{Department of Physics and State Key Laboratory of Low Dimensional Quantum Physics, Tsinghua University, 100084, Beijing, China.}
\affiliation{Beijing Key Laboratory of Cold Atom Quantum Computation, Tsinghua University, 100084, Beijing, China.}
\author{Linghui Chen}
\email{lhchen@iflytek.com}
\affiliation{iFLYTEK Research, Hefei, 230088, China.}
\affiliation{Intelligent Quantum Inception Co., Ltd., Haidian, Beijing, 100083, China.} 
\affiliation{Beijing Key Laboratory of Cold Atom Quantum Computation, Tsinghua University, 100084, Beijing, China.}
\date{\today}

\begin{abstract}

Qubit loss is a major source of error in atom-array quantum computing and invalidates the algebraic structure of the standard stabilizer formalism for quantum error-correcting codes. On the one hand, it complicates decoding; on the other hand, it introduces stochastic flicker patterns in stabilizers as a hallmark of qubit loss. Here, we develop an artificial-intelligence-enabled decoder based on a spatiotemporal graph neural network (STGNN) architecture to extract spatial and temporal correlations from syndrome histories. Our decoder performs a dual-head task, simultaneously correcting standard Pauli errors and identifying the locations of qubit loss. Our decoder achieves significantly higher logical accuracy than both the traditional minimum-weight perfect matching (MWPM) algorithm and even delayed-erasure MWPM decoders that use qubit-loss information from the final round as input. Furthermore, we integrate the decoder into a sliding-window framework for dynamic qubit replenishment. We emphasize that our approach does not involve any additional physical measurements beyond stabilizer measurements, and detects qubit loss and determines its replacement strategy through a purely software-based AI policy. We demonstrate that our approach bounds the accumulation of lost qubits and fully restores the fault-tolerant scaling behavior. For realistic near-term error rates, this work provides a robust and scalable framework for correcting qubit-loss errors, paving the way for more efficient fault-tolerant quantum computation.

\end{abstract}

\maketitle

\section{Introduction}
Currently, the advantage of quantum computation in solving realistic problems is hindered by errors from various sources, including decoherence, gate imperfections, and qubit loss~\cite{preskill2018quantum}. To address these challenges, quantum error correction (QEC) has been developed to detect and correct errors in quantum systems~\cite{shor1995scheme, calderbank1996good, nielsen2010quantum, terhal2015quantum, campbell2017roads}. The most widely used framework is that of stabilizer codes~\cite{gottesman1997stabilizer}, which provide a means to encode quantum information without directly measuring the quantum state itself, but rather by measuring stabilizer operators. Stabilizer codes have been successfully implemented in various physical systems, including superconducting qubits~\cite{Hu2019, google2023suppressing, google2025quantum, putterman2025hardware}, trapped ions~\cite{postler2022demonstration, hong2024entangling, paetznick2024demonstration}, and atom arrays~\cite{bluvstein2024logical, bluvstein2025fault}, demonstrating their effectiveness in protecting quantum information. Although stabilizer codes are highly effective at correcting Pauli-type errors, atom loss does not naturally fit into the standard Pauli-error decoding framework. At the level of stabilizer readout, a lost data qubit can masquerade as a sequence of random Pauli faults, producing single-round syndrome patterns that are difficult to distinguish from those arising from ordinary Pauli noise. Nevertheless, the accumulation of lost qubits degrades the connectivity of the stabilizer code, which directly leads to the failure of logical information preservation. There is a pressing need for a decoder that can simultaneously correct Pauli errors and identify qubit loss events, enabling the reinitialization of lost qubits and the continuation of quantum computation.

\begin{figure*}
    \begin{center}  
    \includegraphics[width=2\columnwidth]{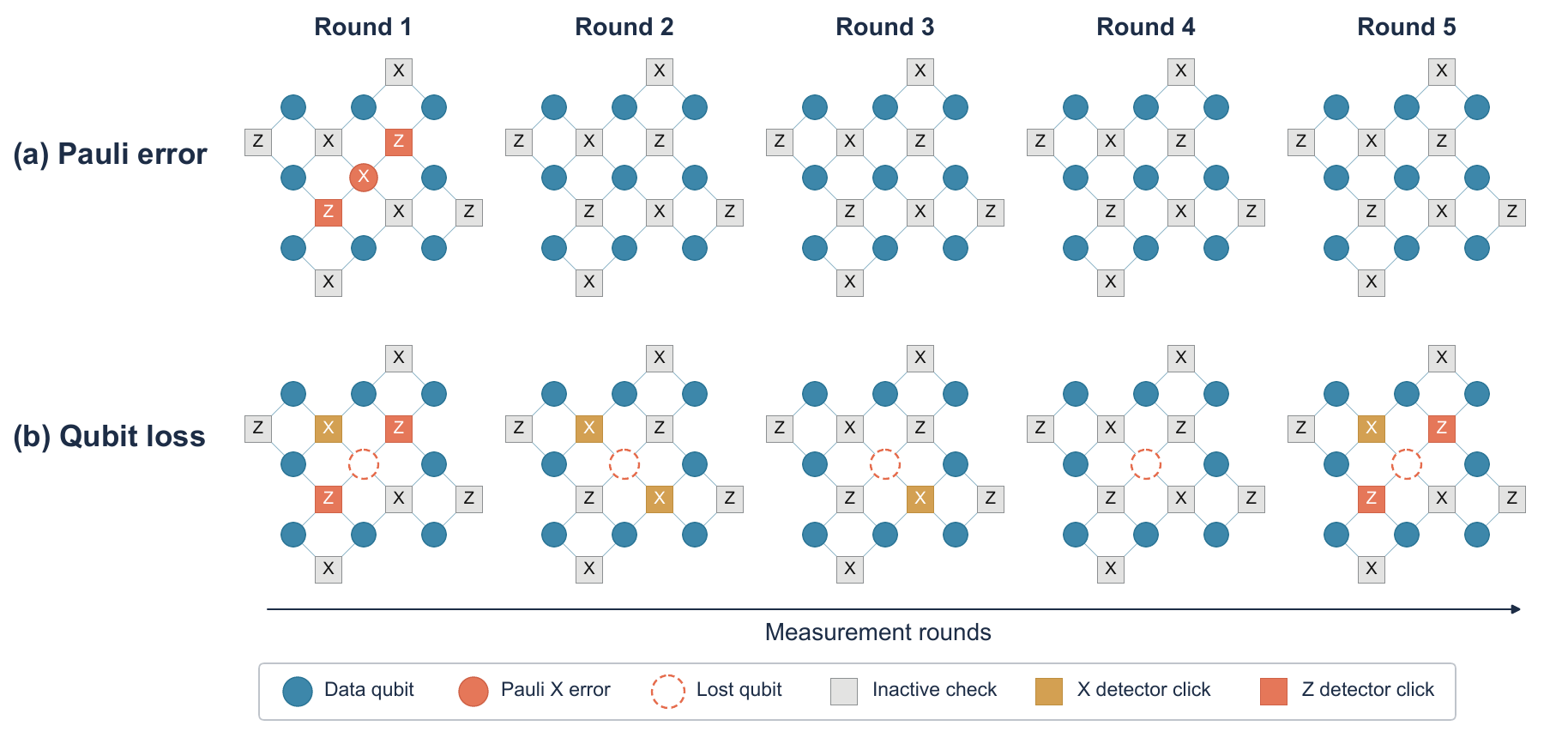}
    \caption{Performance of Pauli error vs. qubit loss in stabilizer measurements using the rotated surface code as an example. A ``Check ON" signal (red or yellow square) represents a detector click, which is defined as a change in the stabilizer measurement value between the current and last round. To mitigate hook errors, the measurement sequence around each data qubit follows a specific order: (1) bottom-right X, (2) bottom-left Z, (3) top-right Z, and finally (4) top-left X. Therefore, diagonally adjacent Z-detectors commute and trigger simultaneously within the same round, whereas the top-left X-detector in each round commutes with the bottom-right X-detector in the next round. (a) Syndrome pattern of a single Pauli $X$ error event. The neighboring $Z$-stabilizers trigger detector clicks (red), while unaffected detectors remain inactive (gray). (b) Syndrome pattern of a qubit loss event. The neighboring $X$- and $Z$-detectors exhibit non-deterministic flickering between active (yellow/red) and inactive (gray) states across multiple rounds, reflecting the stochastic and persistent signature of qubit loss.}
    \label{flicker}
    \end{center}
\end{figure*}

Decoding qubit-loss errors is particularly important in atom array quantum processors. Due to the finite vacuum lifetime, there is an inevitable loss of atomic qubits, which prevents the realization of deep circuits. Although continuous loading techniques have been developed to replenish atomic qubits~\cite{chiu2025continuous}, correcting qubit-loss errors still requires awareness of when and where a qubit is lost. In this regard, alkaline-earth atoms have the advantage that their complex level structure enables detection of atom loss~\cite{scholl2023erasure, ma2023high}, although it also makes quantum control more challenging. For alkali atoms like rubidium, the most popular atomic species for atom array quantum computation, several experimental techniques have also been proposed to detect qubit loss, such as the SWAP-based syndrome extraction (SWAP-SE) technique~\cite{ghosh2013understanding, suchara2015leakage, perrin2025quantum, chow2024circuit, baranes2026leveraging, computing2026quantum, zchg-x177} or logical qubit teleportation~\cite{baranes2026leveraging}. However, these hardware-based protocols are highly invasive: they consume additional experimental resources, require complex path planning, and even inject extra gate noise (Pauli errors) into the system through intermediate probing operations. Hence, it is desirable to decode atom-loss information directly from stabilizer codes without resorting to more complex atomic species or detection techniques.

This is, in principle, possible once the decoder captures correlations across rounds rather than focusing on each round individually. Under the stabilizer-code framework, when a standard Pauli error, such as an $X$ flip, occurs on a physical qubit, it anticommutes with nearby $Z$-type stabilizers, deterministically flipping the signs of those stabilizer outcomes relative to the no-error case, thereby producing a discrete and predictable syndrome pattern. In contrast, when a data qubit is lost, its degree of freedom is removed from the code space; the $X$- and $Z$-type stabilizers that previously overlapped on that qubit become effectively anticommuting in its neighborhood. For instance, if an $X$-stabilizer checks $X_1X_2X_3X_4$ and a $Z$-stabilizer checks $Z_1Z_2Z_5Z_6$, they overlap at an even number of qubits. However, if the first qubit is lost, they become $X_2X_3X_4$ and $Z_2Z_5Z_6$, which overlap on an odd number of qubits and no longer commute. As a result, measuring one type of such stabilizer projects the state onto its $\pm 1$ eigenspace and disturbs the other type; the affected stabilizer outcomes collapse stochastically to $+1$ or $-1$ with equal probability. As shown in Fig.~\ref{flicker}, this stochastic behavior persists across successive measurement rounds as long as the qubit remains missing, yielding temporally correlated ``flickering'' syndromes rather than a single, isolated sign flip~\cite{ghosh2013understanding, varbanov2020leakage}.

Therefore, to properly decode the qubit-loss error, it is crucial to capture the temporal correlation between syndrome measurements across different rounds. Since the probability of Pauli errors mimicking the statistical signature of multiple-round detector flicker is exponentially low, if the decoder collects syndrome information over a sequence of rounds before making a final decision, it can effectively differentiate between Pauli and qubit loss errors. 

Consequently, it is natural to develop a decoder that can simultaneously perform a dual task: decode logical flip and locate qubit-loss errors, a capability that has remained absent until now. Artificial-intelligence (AI)-based decoders have demonstrated superior performance in decoding various codes and learning complex syndrome patterns, surpassing traditional decoding algorithms in accuracy and efficiency~\cite{torlai2017neural, varsamopoulos2017decoding, bausch2024learning, senior2025scalable, hu2025efficient, maan2025machine, blue2025machine, cao2025generative, zhang2026learning}. In this work, we develop an AI decoder that enables the simultaneous capture of spatial and temporal correlations, thereby fulfilling this dual task.

\begin{figure*}
    \begin{center}  
    \includegraphics[width=2\columnwidth]{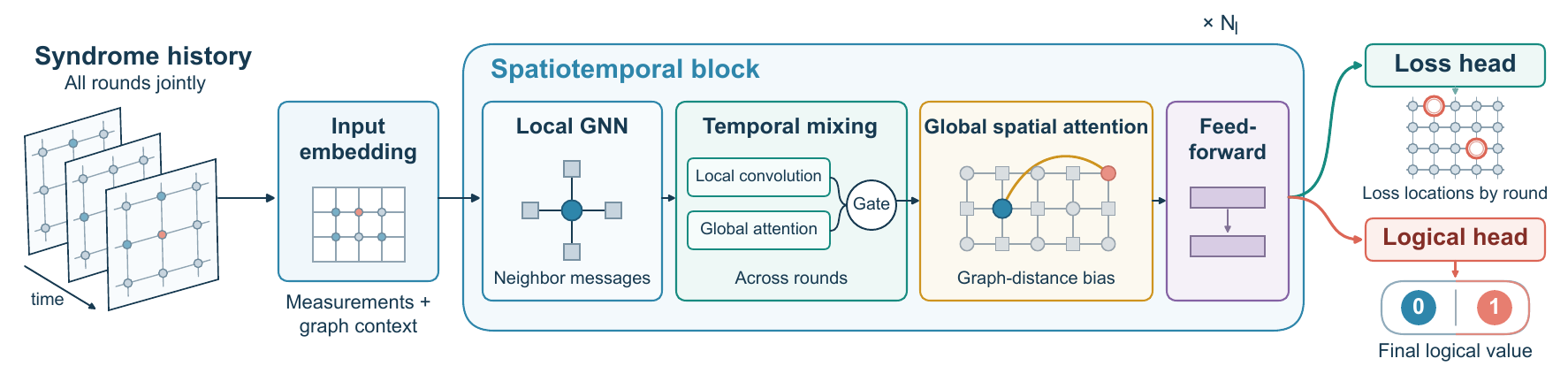}
    \caption{Architectural data flow of the STGNN decoder. The network jointly processes the entire spatiotemporal syndrome volume rather than sequentially. After the initial input embedding, the features pass through $N_l$ spatiotemporal blocks, each consisting of a local GNN, a gated temporal mixing module (fusing local and global temporal correlations), a graph-aware global spatial attention module, and a feed-forward network. The refined features are then passed to dual decoding heads: a loss head for identifying qubit-loss events and a logical head for predicting the final logical value.}
    \label{architecture}
    \end{center}
\end{figure*}

\section{Architecture of the AI Decoder} 
The architecture of our AI decoder is a custom-designed Spatiotemporal Graph Neural Network (STGNN) featuring an interleaved processing structure in the spatial and temporal dimensions, as illustrated in Fig.~\ref{architecture}. The input to the network consists of six primary features for each physical node: (1) the binary ancilla measurement result, (2) the detector results (the XOR difference between consecutive rounds), (3) the node type (data or ancilla), (4) the ancilla type (\textit{X} or \textit{Z}), (5) the task type (X-basis or Z-basis memory), and (6) the node index. Crucially, the complete syndrome history of $T+1$ rounds is fed into the model simultaneously as a single spatiotemporal volume, rather than in a recurrent, round-by-round fashion. In the embedding stage, these features are projected into a high-dimensional hidden tensor $h\in \mathbb{R}^{B\times (T+1)\times N\times D}$ with dimensions for batch size (B), syndrome rounds (T), node count (N), and hidden dimension (D), which serves as the initial state for the subsequent layers. 

The core of the STGNN comprises four specialized sub-layers that progressively refine the error features within each spatiotemporal block (Fig.~\ref{architecture}). First, to facilitate information exchange between data nodes and ancilla nodes, a local GNN performs message passing restricted to the physical connectivity (Tanner graph) of the qubits. Second, to resolve the stochastic ``detector flicker" and capture temporal correlations, a temporal mixing module decouples features into a one-dimensional convolutional branch for short-range transients and a Multi-Head Attention (MHA) branch for long-range global dependencies. These signals are fused via a learnable gating mechanism that adaptively weights the local and global contributions before a residual update. Third, to capture global spatial correlations that exceed the GNN's local receptive field, a global spatial attention module modulates attention scores using a graph-distance bias derived from shortest-path graph distances. Finally, a point-wise feed-forward network (FFN) applies nonlinear projections for further feature refinement.

We repeat this spatiotemporal block $N_{l}$ times, after which the refined spatiotemporal volume is passed to dual decoding heads: a loss head that identifies the round-wise qubit loss coordinates and a logical head that determines the final logical value, providing a comprehensive inference of the system's error status.

We emphasize that the STGNN is not the only structure that can be applied in the qubit loss decoding task. Other architectures, such as the AlphaQubit decoder~\cite{bausch2024learning, senior2025scalable}, can be adapted for this purpose by incorporating an auxiliary qubit-loss head at each measurement round. The fundamental requirement is that the network must be capable of extracting temporal correlation from the syndrome history, as this is the hallmark of qubit loss. However, the sequential RNN-based architecture of such models typically excels at tracking deterministic mean values (e.g., Pauli error trajectories) rather than extracting the temporal covariance inherent in stochastic flicker signatures. By processing the entire syndrome window simultaneously as a unified spatiotemporal volume, the STGNN provides a more direct inductive bias for learning these complex temporal correlations.

\section{Part I: Single-Shot Inference and Offline Validation} 

To prove that our non-invasive STGNN can genuinely decode qubit loss from syndrome flickers, we first evaluate its performance under a static, single-shot inference protocol. In this regime, the decoder processes varying lengths of syndrome history $T$ simultaneously using temporal masking to make a single final logical prediction without intermediate hardware replenishment.

\textit{Offline Error Model and Data Generation.} We train and evaluate the STGNN decoder using a rotated surface code memory task~\cite{bravyi1998quantum, dennis2002topological, fowler2012surface} under a realistic circuit-level noise model. The training data is generated offline via the Stim toolbox~\cite{gidney2021stim}, which enables fast simulation of quantum stabilizer circuits. Each sample consists of a syndrome history with two labels: binary final value of the logical operators, and a binary mask representing the spatiotemporal coordinates of all qubit-loss events. 

Our standard offline error model incorporates single-qubit depolarizing noise (with probability $P_\mathrm{Pauli}$) acting on corresponding qubits after H gate operations, two-qubit depolarizing noise (with probability $P_{\mathrm{CNOT}}$) after CNOT gate operations, and measurement errors (with probability $P_\mathrm{meas}$) during readout. The qubit loss events are modeled as a stochastic process where each physical qubit has a probability $P_\mathrm{loss}$ of being removed in each round. Notably, our noise model distinguishes between ancilla and data-qubit loss. Because ancilla qubits are measured at the end of each round and refreshed at the beginning of the next round, ancilla loss is transient and only affects the measurement outcomes of the current round. In contrast, data-qubit loss is persistent: once a data qubit is lost, it no longer participates in subsequent gate operations, and its effects accumulate over multiple rounds until the decoder identifies the loss and hardware-level reinitialization is performed. Lost data qubits are recorded as being in the $|0\rangle$ state, as the readout process in the atom array platform specifically detects the population of the $|1\rangle$ state, making a loss event indistinguishable from the $|0\rangle$ state in a single measurement. For this study, we set $P_\mathrm{Pauli}=P_\mathrm{CNOT}=P_\mathrm{meas}=0.005$ and $P_\mathrm{loss}=0.015$ to simulate a regime where qubit loss significantly impacts operation continuity and logical accuracy.

\textit{Training process.} The decoders are trained within a supervised learning framework. To optimize the models for both logical prediction and loss-event identification, we employ a multi-task loss function that combines the classification loss (e.g., Cross-Entropy) for logical prediction and loss-event identification. This objective function incorporates tunable weighting coefficients to balance the convergence rates of the two tasks and prevent one from dominating the optimization process.

\begin{figure}
    \begin{center}  
    \includegraphics[width=1\columnwidth]{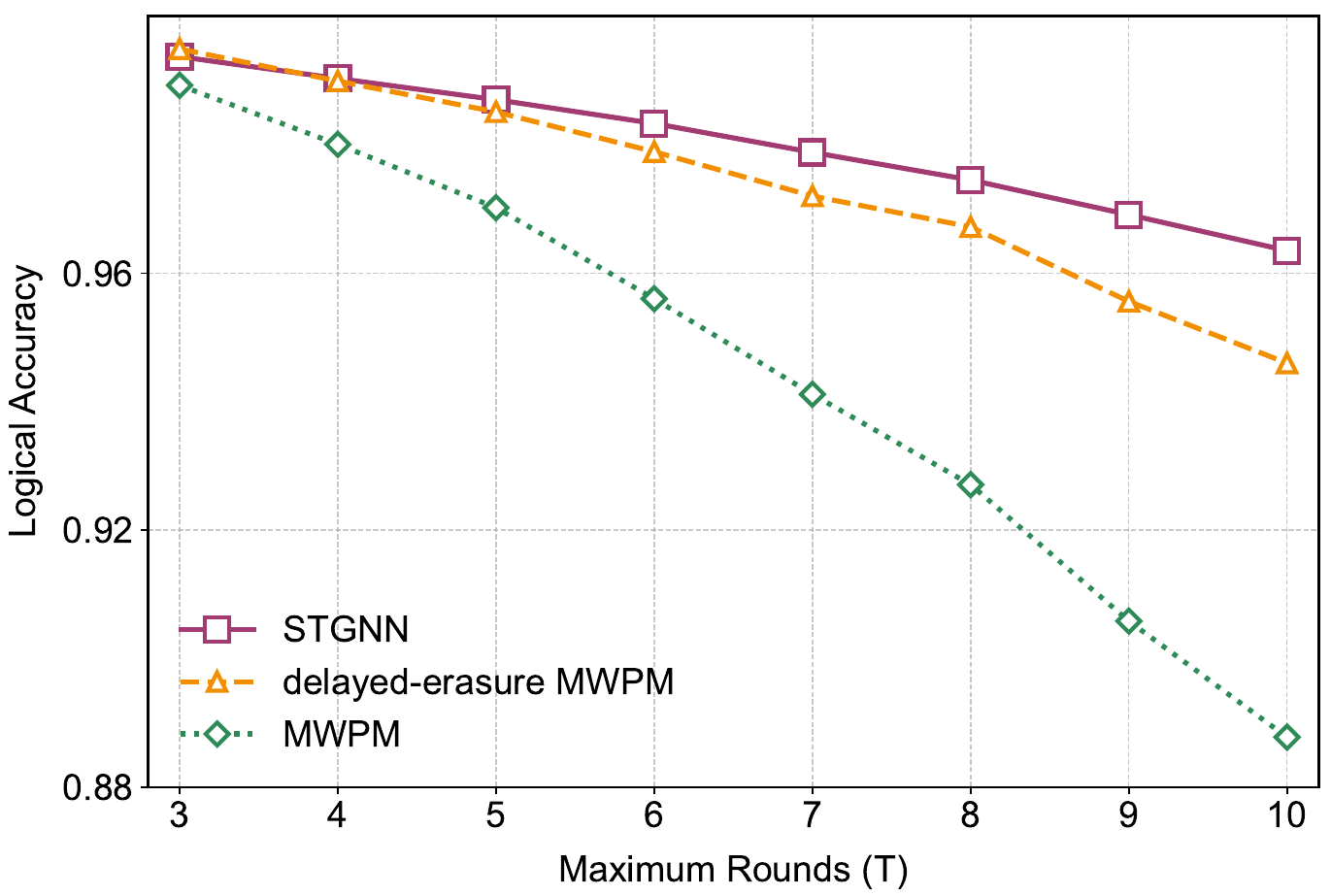}
    \caption{Comparison of decoder performance under qubit loss. Logical accuracy is plotted against the maximum number of syndrome measurement rounds (T) for a rotated surface code with distance $d=5$. STGNN surpasses the standard MWPM decoder and the modified MWPM decoder with final loss information as input.}
    \label{LogicalPerformance}
    \end{center}
\end{figure}

\textit{Logic Accuracy.} We evaluate the performance of the STGNN decoder by benchmarking their logical accuracy and loss-identification metrics. To provide a statistically robust assessment, we define logical accuracy by considering a complete set of $d$ equivalent logical observables (e.g., all $d$ rows or columns across the surface code lattice). If any physical qubit within an operator's support is lost during the QEC rounds, that specific operator is excluded from the accuracy calculation. The final logical accuracy is then determined by the average success rate across all remaining loss-free logical observables. This methodology ensures that the benchmark accurately reflects the decoder's success in preserving quantum information within the surviving code space. 

To demonstrate the superior capability of the STGNN in extracting spatiotemporal correlations, we compare it against two decoding strategies: the standard Minimum Weight Perfect Matching (MWPM) decoder~\cite{dennis2002topological}, and a delayed-erasure MWPM decoder~\cite{baranes2026leveraging}. The delayed-erasure MWPM is provided with the precise spatial locations of lost qubits at the final round, serving as a privileged baseline with access to more physical information than the other decoders.

Fig.~\ref{LogicalPerformance} illustrates the logical accuracy as a function of the number of QEC rounds for a rotated surface code with distance $d=5$. The STGNN significantly outperforms both traditional baselines, exhibiting superior logical fidelity and enhanced scalability with increasing numbers of rounds. Remarkably, the STGNN decoder achieves higher accuracy than even the delayed-erasure MWPM, despite the latter possessing a priori knowledge of the loss locations. This demonstrates that our AI model resolves the spatiotemporal ambiguity of loss events more effectively by extracting correlations from the entire syndrome history, utilizing the temporal flicker as a richer diagnostic tool than just the final state.

\textit{Qubit Loss Identification.} Another critical metric for our decoder is its diagnostic capability in identifying lost data qubits. After 10 rounds of QEC, the STGNN decoder achieves a recall of $0.732$, meaning that $73.2\%$ of lost qubits are correctly identified and flagged for replacement. The precision at the final round is $0.874$, indicating that $12.6\%$ of the qubits predicted as ``lost" were actually still present in the system. While these false positives might initially seem disadvantageous, they often occur in regions where a qubit exhibits high-frequency Pauli errors that mimic the statistical signature of a loss-induced flicker. In such cases, treating these ``noisy" qubits as lost and reinitializing them can effectively reduce the local entropy and prevent error propagation, thereby benefiting the long-term stability of a deep QEC circuit. It should be noted that both the standard and the delayed-erasure MWPM decoders are excluded from this diagnostic comparison, as they lack an inherent loss-identification mechanism and focus exclusively on error correction rather than loss inference.

\begin{figure}
    \begin{center}  
    \includegraphics[width=1\columnwidth]{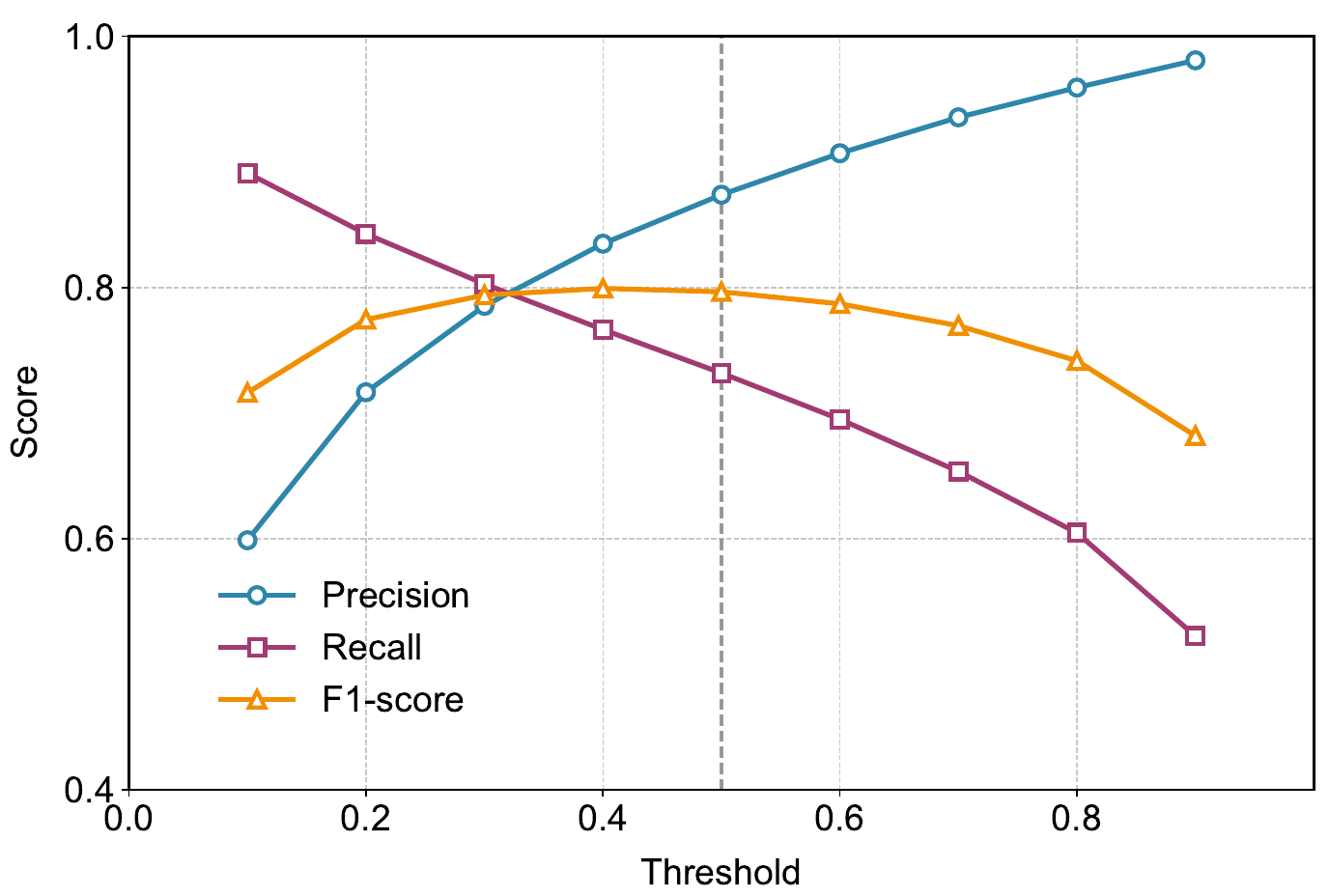}
    \caption{Qubit loss identification performance vs. decision threshold. Precision, Recall, and F1-score are evaluated for the STGNN decoder. The curves demonstrate the decoder's sensitivity to the threshold and its capacity for optimized loss inference.}
    \label{LossThreshold}
    \end{center}
\end{figure}

Specifically, the output of the data-loss head is a predicted probability between $0$ and $1$. In standard classification tasks, the decision threshold is typically set to 0.5. However, this parameter can be adjusted to navigate the trade-off between recall and precision and to optimize the F1-score (the harmonic mean of the two). Fig.~\ref{LossThreshold} illustrates the sensitivity of these metrics to the classification threshold. As expected, increasing the threshold leads to a decrease in recall but a rise in precision. We observe that the F1-score reaches its maximum at a threshold of approximately 0.4. In practical experimental implementation, this threshold can be dynamically tuned to align with the specific re-initialization policy on the hardware. For instance, an aggressive policy might employ a lower threshold to maximize recall, ensuring that as many lost qubits as possible are replaced at the cost of higher false positives. Conversely, under a conservative policy, one would use a higher threshold to minimize unnecessary reinitialization of healthy qubits. Such flexibility allows the decoder to be tailored to the specific error budgets and operational constraints of different platforms.

\begin{figure}[t]
    \begin{center}  
    \includegraphics[width=1\columnwidth]{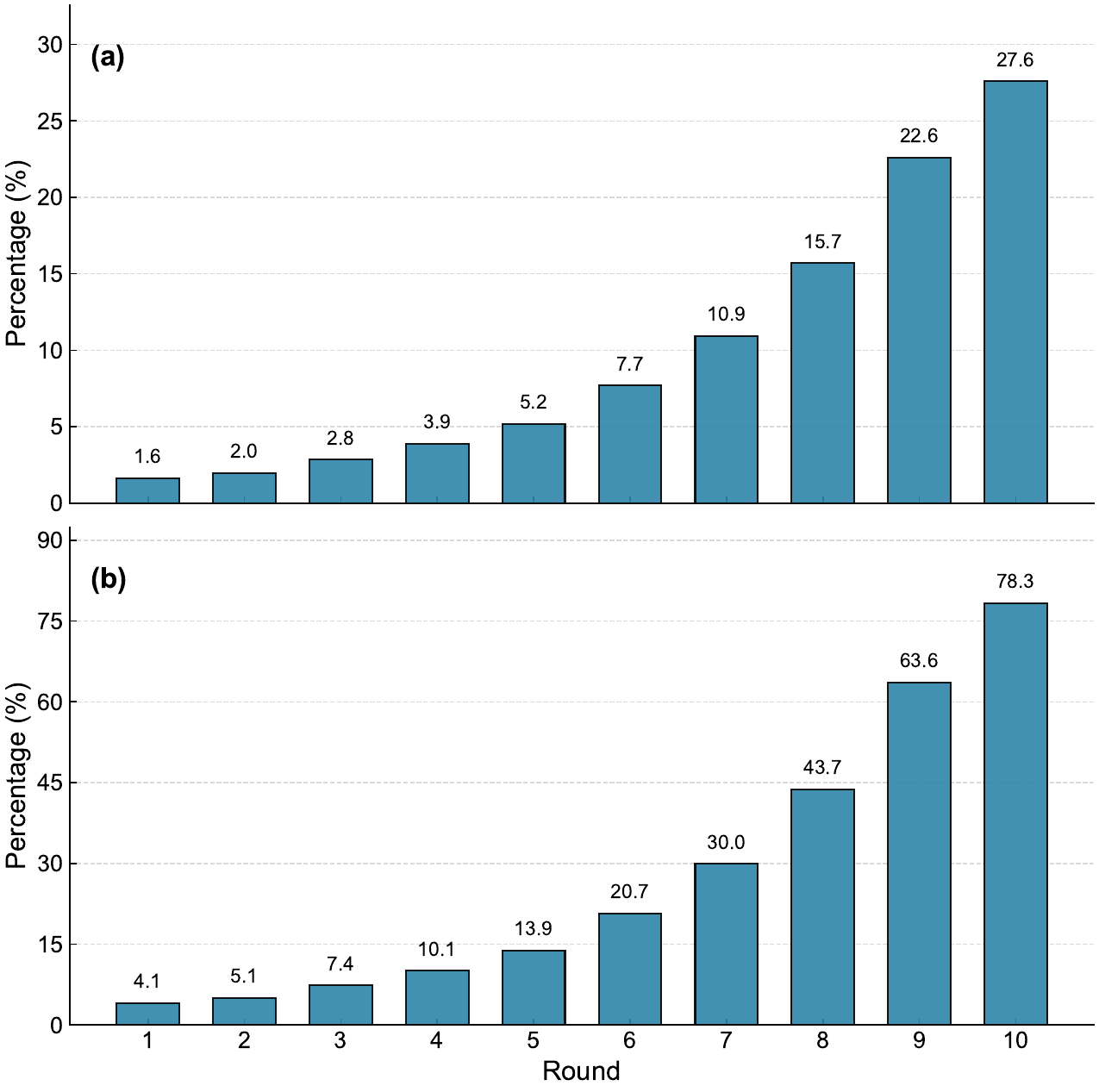}
    \caption{Performance analysis of unrecognized qubit loss after ten QEC rounds. (a) Temporal distribution of False Negatives (FN). The bars represent the percentage of lost qubits not identified by the decoder, categorized by the round in which the loss initially occurred. The majority of missed detections are concentrated in the final few rounds. (b) Miss rate per round. The bars show the proportion of newly occurring loss events in each specific round that remain unrecognized at the final round. As the loss event happens closer to the final round, the likelihood of it being recognized decreases for AI decoders, due to insufficient syndrome history to establish the flicker pattern.}
    \label{LossPerformance}
    \end{center}
\end{figure}

Finally, we analyze the distribution of unrecognized loss events at the final round with a threshold set to 0.5, as shown in Fig.~\ref{LossPerformance}(a). We observe that failures in identifying lost qubits are primarily due to the temporal latency and boundary effects. Qubits lost within the last few rounds do not persist long enough to exhibit the flicker pattern required for high-confidence identification. Among all loss events that remain unrecognized by the final round, only about 1.6\% originate from the earliest round. This latency effect is further quantified in Fig.~\ref{LossPerformance}(b), which illustrates the miss rate relative to the round of occurrence. For losses occurring in the first round, the decoders achieve a successful identification rate of over 95\% (a miss rate $\textless 4.1\%$) due to the sufficient observation window. However, for losses occurring in the final round, the miss rate exceeds 78.3\% for both decoders, as a single round of syndrome data provides negligible statistical evidence to distinguish loss from transient Pauli noise. 

\section{Part II: Deep-Circuit QEC with Dynamic Replenishment}

While the previous section demonstrates the STGNN's superior offline capability in extracting spatiotemporal correlations, a practical fault-tolerant quantum computer sustains deep circuits over extended rounds. In this regime, lost qubits need to be identified and replenished in real time to maintain the integrity of the code structure. 

\begin{figure}[t]
    \begin{center}  
    \includegraphics[width=1\columnwidth]{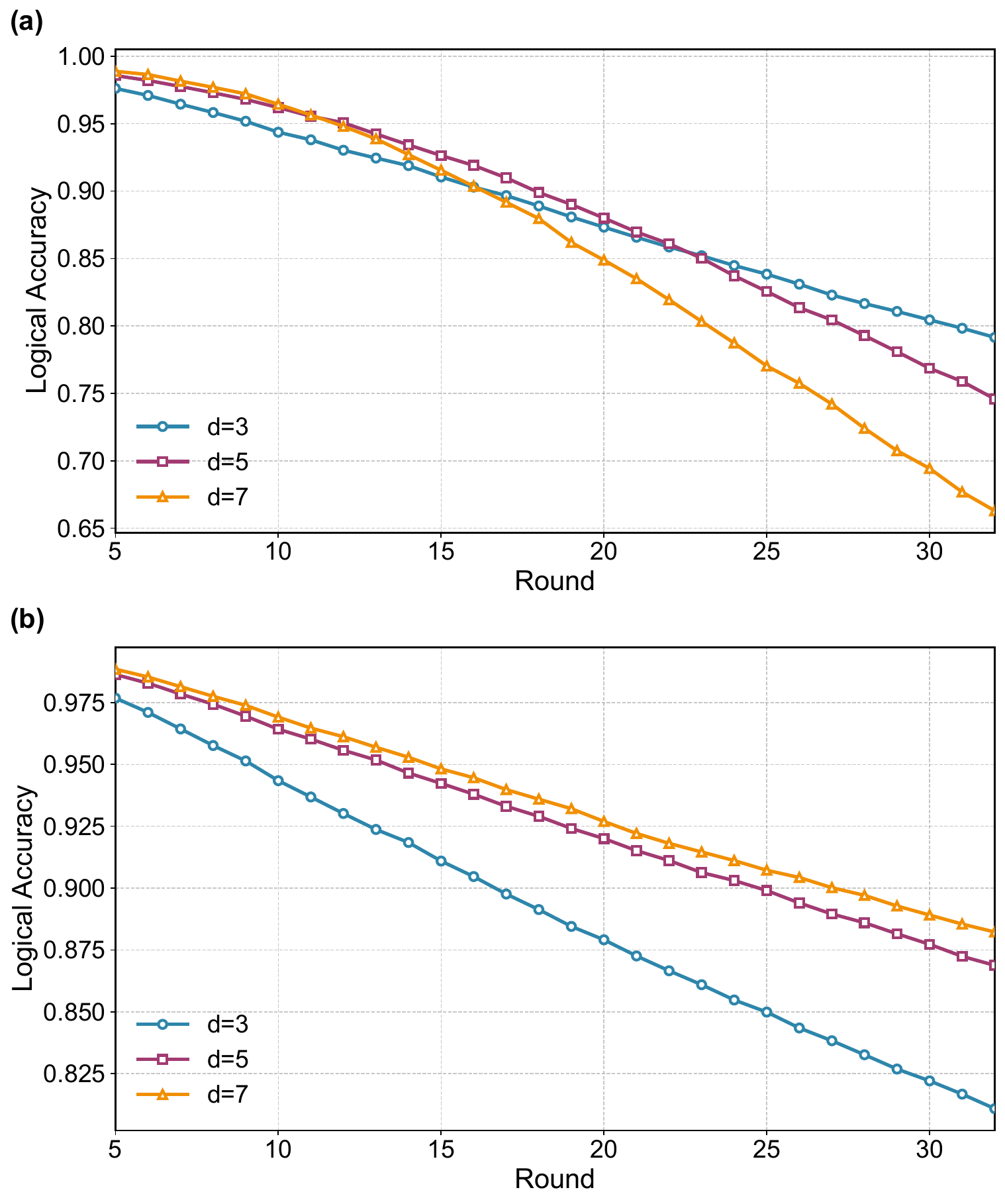}
    \caption{\textbf{Logical accuracy in deep-circuit quantum error correction with and without dynamic qubit replenishment.} (a) Without replenishment, larger code distances exhibit faster logical degradation over extended rounds due to the accumulation of lost qubits. (b) With the proposed AI-based replenishment policy, the expected positive scaling behavior is restored, with larger code distances achieving higher logical accuracy.}
    \label{continuous_distance}
    \end{center}
\end{figure}

Fig.~\ref{continuous_distance}(a) illustrates the logical accuracy of surface codes with different code distances under extended rounds without replenishment. In the early rounds, where Pauli errors dominate, increasing the code distance improves logical protection. However, as the circuit runs deeper, unreplenished qubit losses accumulate and gradually become the dominant failure mechanism. From the perspective of percolation theory~\cite{stace2009thresholds,colmenarez2025fundamental}, once the accumulated losses push the effective loss rate above the critical percolation threshold, larger code distances experience a sharper drop in intact logical connectivity, leading to more severe degradation of the encoded logical information. This inverted scaling behavior highlights the necessity of dynamic qubit replenishment for multi-round fault-tolerant operation.


\textit{The Sliding-Window Protocol.} Extending the full-sequence inference of Part I directly to deep circuits is impractical, as the growing history size increases computational overhead and cannot provide real-time, mid-circuit feedback for qubit replenishment. To enable dynamic replenishment during extended rounds, we deploy the same STGNN architecture within a sliding-window protocol~\cite{dennis2002topological, skoric2023parallel, tan2023scalable, zhang2026learning}. Instead of processing the entire circuit history at once, the network is adapted to operate on a fixed-length rolling window of syndrome history (specifically, 5 consecutive rounds) and additionally incorporates recent replenishment actions into its data-node features. At each step, the decoder outputs loss predictions for the current window. Any qubit flagged as lost is physically replenished and reinitialized before the next round. The window then slides forward by one round (stride 1), allowing the physical qubit count to be dynamically maintained. 

In our simulations, we assume idealized error-free replenishment, where flagged data qubits are instantaneously reinitialized to the $|0\rangle$ state. This abstraction allows us to isolate and benchmark the intrinsic performance of the software-based loss decoder. In practice, finite reloading infidelities can be absorbed into the effective physical error rates.

\begin{figure}[t]
    \begin{center}  
    \includegraphics[width=1\columnwidth]{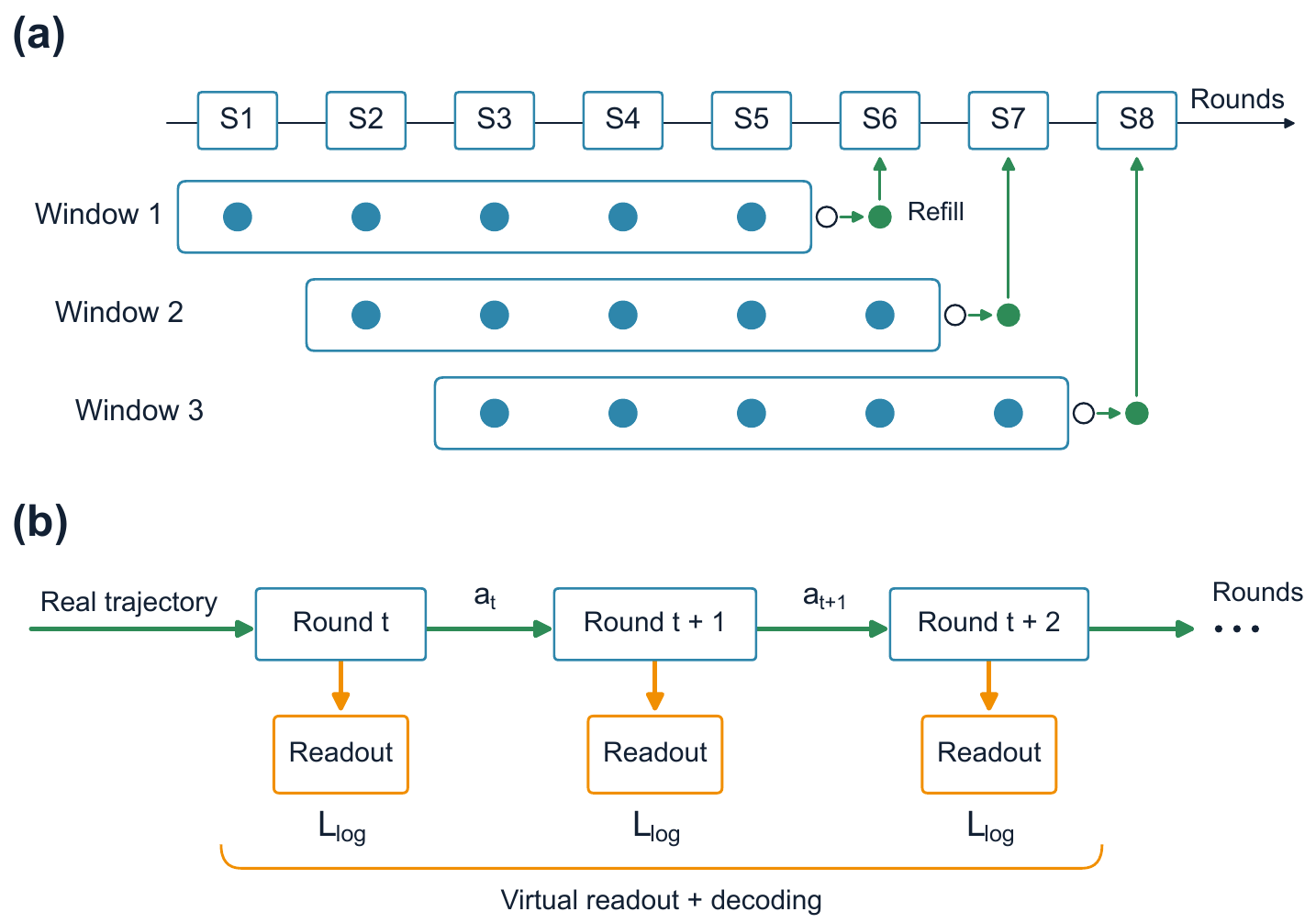}
    \caption{\textbf{The fishbone training strategy for the online sliding-window protocol.} (a) Sliding-window inference. For extended multi-round operation, the decoder processes a fixed history of syndrome rounds $S_t$ (e.g., a window of 5 rounds) to predict qubit loss locations (followed by qubit replenishment, labeled ``Refill''). The window slides forward with a stride of 1, allowing for real-time identification and replenishment. (b) Fishbone training. To provide dense supervision in an online environment, the training process splits into two branches at each window. The real branch executes replenishment actions $a_t$ and is supervised by ground-truth loss labels, while the virtual branch attaches a hypothetical ideal final readout to compute logical logits $L_{\mathrm{log}}$, providing dense logical supervision without interrupting the continuing physical trajectory.}
    \label{fishbone_training}
    \end{center}
\end{figure}

\textit{Online Training Strategy.} 
In the dynamic replenishment regime, training data must be generated interactively since the circuit connectivity dynamically changes based on the decoder's replenishment decisions. To train the AI decoder effectively, we employ a multi-stage strategy starting with online supervised learning (SL). We introduce a ``fishbone'' training structure (Fig.~\ref{fishbone_training}) that attaches virtual final-round branches after each sliding window to provide dense logical supervisory signals. Online SL provides a stable and effective initialization for the AI decoder, allowing it to learn how qubit loss manifests as syndrome flickers while simultaneously learning to decode logical information.

As discussed in Part I, the classification threshold governs the trade-off between false positives and delayed detections. To locate the optimal operating point, we scan the threshold post-training to maximize the final logical accuracy. The optimal threshold is then employed to evaluate the long-term stability of the decoder in a multi-round QEC setting.

\textit{Results.} We evaluate the long-term performance of the AI policy over 32 QEC rounds across different code distances. The AI policy successfully identifies and replenishes lost qubits, maintaining a stable physical qubit survival fraction of approximately 96.5\% for $d=5$ throughout extended QEC rounds. This demonstrates that the proposed software-based decoder effectively bounds the accumulation of physical defects during multi-round operation. 

The ultimate test of any error-correction scheme is whether it restores the fault-tolerant scaling behavior of the code. In Fig.~\ref{continuous_distance}(a), we demonstrated a pathological crossing phenomenon where larger code distances suffer faster logical degradation due to the unmitigated accumulation of lost qubits. To verify that our approach resolves this, we evaluate the AI policy across varying code distances ($d=3,5,7$) in Fig.~\ref{continuous_distance}(b). The logical accuracy now improves monotonically with increasing code distance. The restoration of the positive scaling behavior confirms that our non-invasive AI policy successfully protects the structural integrity of the code across different sizes, providing a scalable pathway for large-scale quantum computation, even in the presence of significant qubit loss.

\section{Comparison with Hardware-based Loss Detection}

While our work proposes a purely software-driven AI policy to infer qubit loss from syndrome histories, alternative hardware-based protocols, such as the SWAP-SE technique~\cite{ghosh2013understanding, suchara2015leakage, perrin2025quantum, chow2024circuit, baranes2026leveraging, computing2026quantum, zchg-x177}, have been proposed for active loss detection. In a SWAP-SE protocol, the roles of physical data and ancilla qubits are periodically exchanged, followed by active detection using State-Selective Readout (SSR)~\cite{baranes2026leveraging}. In this section, we qualitatively compare the two approaches, highlighting their respective advantages and limitations. The comparison between the two approaches is summarized in Table~\ref{tab:comparison}.

\textit{Latency vs. Invasiveness.} The primary advantage of the SWAP-SE protocol is its low detection latency. A lost data qubit can be identified almost immediately after the swapping and state-selective readout operations. However, this rapid detection comes at the cost of additional physical operations. The extra SWAP gates and intermediate SSR measurements introduce additional gate and measurement errors into the system. In contrast, our AI policy is completely non-invasive, relying exclusively on standard stabilizer measurement outcomes without introducing additional quantum operations. Its main limitation is an intrinsic algorithmic latency, as approximately 3-6 rounds of syndrome history are required to accumulate sufficient statistical evidence to distinguish loss-induced flickering from transient Pauli errors. Nevertheless, as demonstrated in the previous section, the AI policy maintains a stable physical survival fraction during multi-round operation, allowing the code to tolerate this identification delay while preserving its connectivity integrity.

\textit{Layout Compatibility and Routing Overhead.} A critical, often overlooked limitation of the SWAP-SE protocol is its rigid spatial-layout requirements. Efficient implementation of the SWAP-SE protocol benefits from a one-to-one correspondence between data and ancilla qubits. This is naturally satisfied in a toric code with periodic boundary conditions~\cite{computing2026quantum}. However, executing a SWAP-SE protocol on a surface code requires complex physical routing, because of the unequal number of data and ancilla qubits. In contrast, the STGNN decoder learns the correlations directly from the Tanner graph, and naturally adapts to the boundary truncation of the surface code without requiring any changes to the physical qubit layout or scheduling.


\begin{table*}[!t]
\centering
\caption{Qualitative comparison between the hardware-based SWAP-SE protocol and our software-based AI policy for qubit loss detection.}
\label{tab:comparison}
\renewcommand{\arraystretch}{1.4}
\begin{tabular}{p{3.7cm} p{6.5cm} p{6.5cm}}
\hline\hline
\textbf{Feature} & \textbf{Hardware-Based (SWAP-SE)} & \textbf{Software-Based (Our AI Policy)} \\ 
\hline
\textbf{Detection Mechanism} & Active physical swapping and state-selective readout (SSR). & Passive inference from standard spatiotemporal syndrome histories. \\
\textbf{Detection Latency} & \textbf{Low but finite} (Immediate deterministic detection). & $\sim$3-6 rounds (Requires time to accumulate statistical ``flicker" evidence). \\
\textbf{Invasiveness} & \textbf{High} (Injects additional Pauli noise via extra SWAP gates and intermediate measurements). & \textbf{Zero} (Completely non-invasive; no extra physical operations applied to the array). \\
\textbf{Code-Structure Generality} & \textbf{Restricted} (easy for Toric codes, difficult for open-boundary Surface codes). & \textbf{General} (naturally adapts to the Tanner graph of any code structure). \\
\textbf{Hardware Overhead} & Requires complex atom routing, additional lasers for SSR, and sophisticated pulse scheduling. & No extra quantum hardware required; relies entirely on classical neural-network post-processing. \\
\hline\hline
\end{tabular}
\end{table*}







\textit{Conclusion.}
In summary, we have developed an AI-based decoder that can simultaneously decode Pauli and qubit-loss errors directly from stabilizer measurement results, providing a compelling software-driven alternative to complex, invasive hardware detection protocols. This is particularly important and timely for the atom array platform because loss of an atomic qubit is inevitable, and the continuous loading technique is ready to correct this error once the qubit-loss information can be decoded. This capability is part of our broader efforts to build a powerful AI-based decoder, ``Bian Que" \footnote{Bian Que was a famous doctor in ancient China who excelled at diagnosing a person's illness through the non-invasive method of observing, listening, questioning, and pulse-taking, which is very similar to inferring errors through stabilizer measurements.}, for fault-tolerant quantum computation in atom array platforms. Future efforts along this direction include correcting errors in logic gate operations~\cite{zhou2025learning}, building a foundation model with generalization ability, and decoding non-local qLDPC codes~\cite{hu2025efficient, maan2025machine, blue2025machine}.

\begin{acknowledgments}
This work is supported by the National Natural Science Foundation of China under Grant No.~12488301 (H.Z.) and No.~U23A6004 (H.Z.).
\end{acknowledgments}

\bibliography{QEC} 

\newpage
\begin{center}
\textbf{\large Supplemental Material for: AI-Enabled Qubit Loss Detection and Replenishment Strategy for Atom Array Quantum Computing}
\end{center}
\setcounter{equation}{0}
\setcounter{figure}{0}
\renewcommand{\theequation}{S\arabic{equation}}
\renewcommand{\thefigure}{S\arabic{figure}}

\section{Network Structure and Training Details}

\subsection{Architecture of the Spatiotemporal Graph Neural Network (STGNN) Decoder}
\label{app:stgnn_decoder}

The STGNN architecture is a custom-designed framework specifically engineered to capture the distinct signatures of qubit loss within the code's hardware connectivity. Unlike recurrent architectures that process data round-by-round, the STGNN treats the entire syndrome history as a unified spatiotemporal volume, allowing for non-causal feature extraction across the full decoding window.

\subsubsection{Notation}
\label{app:stgnn_notation}

We consider a surface code experiment with $T$ rounds of stabilizer measurements followed by a final destructive readout at $t=T$. Let $\mathcal{V} = \mathcal{A} \cup \mathcal{D}$ denote the set of physical nodes, where $\mathcal{A}$ represents the set of ancilla (stabilizer) qubits and $\mathcal{D}$ represents the set of data qubits. The total number of nodes is $N = |\mathcal{V}|$. The decoder operates on a high-dimensional hidden state tensor $\mathbf{h} \in \mathbb{R}^{B \times (T+1) \times N \times D}$, where $B$ is the batch size, $T$ is the number of syndrome rounds, $N$ is the node count, and $D$ is the hidden dimension.

\subsubsection{Input and Embedding Stage}
\label{app:stgnn_inputs}

The input to the network for each physical node $i \in \mathcal{V}$ at round $t \in \{0, \dots, T\}$ comprises six primary features:
\begin{enumerate}
    \item The binary measurement outcome $m_i^t \in \{0, 1\}$.
    \item The detector result $e_i^t = m_i^t \oplus m_i^{t-1}$ (defined for $t > 0$).
    \item The node type (data qubit or ancilla qubit).
    \item The ancilla type ($X$-type or $Z$-type stabilizer).
    \item The memory task type ($X$-basis or $Z$-basis memory).
    \item The node index $i \in \{1, \dots, N\}$.
\end{enumerate}

During the embedding stage, these features are projected into the latent space and combined with learned embeddings to form the initial hidden state $\mathbf{h}_i^{t, (0)}$:
\begin{equation}
    \mathbf{h}_i^{t, (0)} = \text{Embed}_{\text{feat}}(\mathbf{x}_i^t) + \mathbf{E}_{\text{type}} + \mathbf{E}_{\text{task}} + \mathbf{E}_{\text{index}},
\end{equation}
where $\mathbf{E}_{\text{index}} \in \mathbb{R}^{N \times D}$ allows the network to learn position-specific noise characteristics and boundary conditions across the lattice.

\subsubsection{Interleaved Spatiotemporal Layers}
\label{app:stgnn_layers}

The core of the STGNN consists of $N_{\text{layer}}$ blocks, each comprising three specialized sub-layers arranged in an interleaved sequence. To ensure numerical stability, we employ residual scaling ($\text{ResScale} = 0.1$) for each sub-layer update.

\paragraph{Hetero-Local GNN.} 
This layer facilitates information exchange between data nodes and ancilla nodes. It performs message passing restricted to the Tanner graph edges $\mathcal{E}$. The messages are computed in a heterogeneous manner to distinguish different qubit roles:
\begin{equation}
    \mathbf{m}_{i \leftarrow j} = 
    \begin{cases} 
    \mathbf{W}_{d \to c} \mathbf{h}_j^{t}, & \text{if } j \in \mathcal{D}, i \in \mathcal{A} \\
    \mathbf{W}_{c \to d} \mathbf{h}_j^{t}, & \text{if } j \in \mathcal{A}, i \in \mathcal{D} 
    \end{cases}
\end{equation}
The node state $\mathbf{h}_i^{t}$ is updated as:
\begin{equation}
\begin{aligned}
    \mathbf{h}_i^{t, \text{new}} = \text{LayerNorm} \Bigl( & \mathbf{h}_i^{t} + \alpha \cdot \text{GELU} \bigl( \mathbf{W}_{\text{self}} \mathbf{h}_i^{t} \\
    & + \sum_{j \in \mathcal{N}(i)} \mathbf{m}_{i \leftarrow j} \bigr) \Bigr),
\end{aligned}
\end{equation}
where $\alpha = 0.1$ is the residual scaling factor.

\paragraph{Gated Temporal Mixer.} 
To resolve the stochastic ``detector flicker,'' the temporal mixer decouples features into a local convolutional branch and a global attention branch. For each node $i$, the fusion is governed by a learnable gating mechanism:
\begin{equation}
    g = \sigma\left( \text{Linear}\left( \text{detach}(\mathbf{h}_{\text{conv}}) \right) \right)
\end{equation}
\begin{equation}
    \mathbf{h}_{\text{out}} = g \cdot \mathbf{h}_{\text{conv}} + (1 - g) \cdot \mathbf{h}_{\text{attn}}
\end{equation}
where $\mathbf{h}_{\text{conv}}$ is processed via 1D-convolutions along the temporal dimension $T$, and $\mathbf{h}_{\text{attn}}$ is processed via Multi-Head Attention (MHA). The \textit{detach} operator ensures the gating signal $g$ is learned independently of the convolutional feature gradients.

\paragraph{Global Spatial Attention.} 
This module captures long-range spatial correlations beyond the GNN's local receptive field. Let $\mathcal{D}_{ij}$ be the shortest-path distance between nodes $i$ and $j$ on the Tanner graph. The attention scores are modulated by a graph-distance bias $B$:
\begin{equation}
    \text{Score}_{ij} = \frac{(\mathbf{Q} \mathbf{h}_i)(\mathbf{K} \mathbf{h}_j)^T}{\sqrt{d_k}} + B(\min(\mathcal{D}_{ij}, 24))
\end{equation}
where $B$ is a learned embedding representing the graph-distance prior.

\subsubsection{Decoding Heads}
\label{app:stgnn_heads}

\paragraph{Data Loss Head.} 
A per-node Multi-Layer Perceptron (MLP) is applied to the refined hidden states of data qubits $j \in \mathcal{D}$. It outputs the predicted probability of a qubit loss event $\boldsymbol{\ell}_j^t$ for each node $j$ and each round $t \in \{0, \dots, T\}$.

\paragraph{Logical Head.} 
To predict the final logical state $L$, the network aggregates features from all data qubits $\mathbf{h}_j$ ($j \in \mathcal{D}$). The information is pooled along spatial logical strings using mean and max pooling, followed by 1D-convolutions. Finally, a masked max-pooling operation is performed across the temporal dimension $t$ to ensure the prediction is robust against the variable timing of errors and loss events.

\subsubsection{Loss Function}
Let $y_{j}^t \in \{0,1\}$ denote the binary loss label for data qubit $j$ at round $t$ (with $1$ indicating ``lost''),
and let $y_k \in \{0,1\}$ denote the binary ground-truth logical label for line $k$.
We use a weighted sum of two-class cross-entropies:
\begin{eqnarray}
\mathcal{L} &=&
\lambda_{\mathrm{log}} \sum_{k=1}^{d} \mathrm{CE}\!\left(L_k,\; y_k\right)\nonumber\\
&+&
\lambda_{\mathrm{loss}} \sum_{t=0}^{T}\sum_{j\in\mathcal{D}} \mathrm{CE}\!\left(\boldsymbol{\ell}_j^t,\; y_j^t\right),
\end{eqnarray}
where $\mathrm{CE}(\mathbf{s},y)$ denotes the standard two-class cross-entropy between logits $\mathbf{s}\in\mathbb{R}^2$ and label $y\in\{0,1\}$.
The weights $\lambda_{\mathrm{log}}$ and $\lambda_{\mathrm{loss}}$ are tuned on validation data.

\subsubsection{Implementation Summary and Hyperparameters}
\label{app:stgnn_summary}

The complete forward computation of the STGNN decoder—characterized by the simultaneous processing of $T+1$ syndrome rounds through $N_{\text{layer}}$ interleaved blocks—is synthesized into a unified pipeline. This operational flow encompasses initial embedding, spatiotemporal feature refinement, and dual-head inference for logical state and qubit loss. 

The operational flow of the STGNN forward pass is summarized in Algorithm~1. Specific architectural hyperparameters used in our implementation, including dimensions for the gated temporal mixer and the graph-aware spatial attention, are listed in Table~\ref{tab:stgnn_hparams}.


\vspace{0.4cm}
\noindent\rule{\columnwidth}{0.8pt}

\vspace{-0.1cm}\noindent \textbf{Algorithm 1} STGNN decoder forward pass

\vspace{-0.2cm}\noindent\rule{\columnwidth}{0.4pt}

\begin{algorithmic}[1]
\Require Spatiotemporal binary inputs $\{m_i^t, e_i^t\}_{i \in \mathcal{A}, t=0..T}$ and metadata (node type, task, index).
\Ensure Qubit-loss logits $\{\boldsymbol{\ell}_j^t \in \mathbb{R}^2\}_{j \in \mathcal{D}, t=0..T}$ and logical flip logit $L \in \mathbb{R}^2$.
\State $\mathbf{h}^{(0)} \gets \mathrm{InputEmbedding}(\text{Syndromes}, \text{Metadata})$ \Comment{$\mathbf{h} \in \mathbb{R}^{B \times (T+1) \times N \times D}$}
\For{$l = 1$ to $N_{\text{layer}}$}
  \State $\mathbf{h}_{\text{gnn}} \gets \mathrm{LocalGNN}(\mathbf{h}^{(l-1)})$ \Comment{Heterogeneous spatial message passing}
  \State $\mathbf{h}_{\text{temp}} \gets \mathrm{TemporalMixer}(\mathbf{h}_{\text{gnn}})$ \Comment{Gated fusion of Conv1D and MHA}
  \State $\mathbf{h}_{\text{spat}} \gets \mathrm{GlobalSpatialAttention}(\mathbf{h}_{\text{temp}})$ \Comment{Graph-distance-aware bias}
  \State $\mathbf{h}^{(l)} \gets \text{LayerNorm}(\mathbf{h}^{(l-1)} + \alpha \cdot \mathbf{h}_{\text{spat}})$ \Comment{Residual update with scaling $\alpha = 0.1$}
\EndFor
\State $\{\boldsymbol{\ell}_j^t\}_{j \in \mathcal{D}, t=0..T} \gets \mathrm{Head}_{\text{loss}}(\mathbf{h}^{(N_{\text{layer}})})$
\State $L \gets \mathrm{Head}_{\text{log}}(\mathbf{h}^{(N_{\text{layer}})})$
\end{algorithmic}

\vspace{-0.2cm}\noindent\rule{\columnwidth}{0.8pt}
\vspace{0.4cm}

\begin{table}[h]
\centering
\caption{STGNN decoder architectural hyperparameters.}
\label{tab:stgnn_hparams}
\begin{tabular}{ll}
\hline
\textbf{Component} & \textbf{Value} \\
\hline
Code distance $d$ & 5 \\
Hidden dimension $D$ & 256 \\
Residual scaling factor $\alpha$ & 0.1 \\
Number of interleaved layers $N_{\text{layer}}$ & 6 \\
\hline
\textbf{Temporal Mixer} & \\
\quad Conv1D kernel size & 3 \\
\quad Number of attention heads & 8 \\
\quad Gating mechanism & Sigmoid-activated Linear \\
\hline
\textbf{Spatial Attention} & \\
\quad Maximum graph distance & 24 \\
\quad Distance embedding dimension & 8 \\
\hline
\textbf{Total parameters} & $\sim 8\text{M}$ \\
\hline
\end{tabular}
\end{table}

\section{Error Model and Evaluation Metrics}

\subsection{Circuit-Level Noise Model}
\label{app:circuit_noise}
We validate the decoder using a rotated surface code memory task with code distance $d=5$. The simulation begins by initializing data qubits in a specific basis, followed by $T$ rounds of QEC, and a final ideal data qubit measurement. The noise model includes:
\begin{itemize}
\item \textit{Gate Noise}: Each executed Hadamard is followed by single-qubit depolarization with probability $P_{pauli}=0.005$, and each executed CNOT is followed by two-qubit depolarization with probability $P_{CNOT}=0.005$.
\item \textit{Measurement Noise}: A present ancilla's binary outcome flips with probability $P_{meas}=0.005$. Lost ancilla qubits are treated as state 0.
\item \textit{Qubit Loss}: The nominal loss parameter $P_{loss}=0.015$ is divided among five candidate positions, each with probability $P_{loss}/5=0.003$: round start and the four post-CNOT layers. Post-gate loss applies only to endpoints of executed gates. Lost qubits are removed from the circuit and do not participate in subsequent operations. Data loss persists until replenishment, while ancilla loss is transient and reinitialized in the next round.
\end{itemize}

\subsection{Metrics}
\textit{Logical Accuracy}: We define the logical accuracy as the probability of correctly predicting the logical observable. In cases where a data qubit involved in the logical operator is lost during the QEC rounds, that logical operator is excluded from the accuracy calculation. This approach ensures that the logical accuracy metric accurately reflects the decoder's performance in preserving quantum information within the surviving code space.

\textit{Precision and Recall}: Qubit loss identification is evaluated using precision and recall metrics. Precision is defined as the ratio of true positive loss identifications to the total number of qubits predicted as lost by the decoder. Recall is defined as the ratio of true positive loss identifications to the total number of actual lost qubits. These metrics provide insights into the decoder's ability to accurately identify lost qubits while minimizing false positives. As stated in the main text, the recall is more vital in our QEC framework, as it directly impacts the effectiveness of qubit reinitialization and recovery.

\section{Data Generation, Training, and Evaluation Details}
\label{app:training_evaluation}

We distinguish the full-sequence decoder used for the single-shot inference (Part I) from the sliding-window decoder used for dynamic replenishment (Part II). All experiments are based on rotated-surface-code $Z$-memory tasks. The circuit-level noise and the specific ancilla readout rules are detailed in the ``Circuit-level Noise Model'' section above.

\subsection{System Representation and Supervised Targets}

For both operational modes, the STGNN decoder utilizes the six-dimensional spatiotemporal node features defined in the ``Input and Embedding Stage'' section above. For multi-round operation, data-node features additionally encode the recent hardware replenishment actions over the past three-round histories. Importantly, the ground-truth physical loss mask is never supplied as an input to the decoder.

To evaluate logical fidelity robustly, the logical head predicts the binary parities of $d$ equivalent line representatives of the logical $Z$ observable (e.g., all $d$ columns). A specific representative is included in the logical objective and final accuracy calculation only if every data qubit within its support is physically present at the measurement endpoint. This dynamically masks out ill-defined logical operators whose connectivity has been destroyed by qubit loss. 

\subsection{The Fishbone Training Strategy for Continuous Operation}

A fundamental challenge in training a dynamic replenishment policy is the sparsity of logical supervision, as the true logical state can only be definitively measured at the end of the entire quantum circuit. Furthermore, the circuit connectivity dynamically changes based on the decoder's intermediate replenishment actions. To address this, we implement a Fishbone Training Strategy, which provides dense supervision for both data-loss localization and long-horizon logical accuracy.

The strategy operates on a sliding window of 5 frames with a stride of 1. At each step, the training graph splits into two branches:
\begin{enumerate}
    \item \textbf{The Real Branch (Replenishment Policy):} The decoder processes the current window to predict the data-loss probabilities, from which the actual physical replenishment actions are sampled. This branch is directly supervised by the simulator's ground-truth loss mask. To encourage timely replenishment, we apply a latency-weighted cross-entropy loss $\mathcal{L}_{\mathrm{data,latency}}$, where older unreplenished losses incur exponentially higher penalties (e.g., weight 5 for latency $\geq 6$ vs. weight 0.1 for latency 2).
    
    \item \textbf{The Virtual Branch (Logical Supervision):} To evaluate the logical consequence of the current policy, we construct a hypothetical ``virtual tail.'' Without modifying the continuing physical trajectory, a simulator copy performs a destructive ideal measurement of all data qubits immediately after the current window. This yields the on-basis stabilizers and the logical ground-truth. The decoder processes this virtual tail to compute the focal loss for logical prediction, $\mathcal{L}_{\mathrm{logical}}$.
\end{enumerate}

The overall objective during the online training is a hybrid loss:
\begin{equation}
    \mathcal{L}_{\mathrm{AI}} = \mathcal{L}_{\mathrm{data,latency}} + \lambda_{\mathrm{logical}} \mathcal{L}_{\mathrm{logical}},
\end{equation}
where $\lambda_{\mathrm{logical}}$ balances the immediate physical loss detection with the ultimate goal of preserving logical information.

\subsection{Optimization Pipeline and Threshold Calibration}

The decoders are trained using a curriculum learning approach. The model is first pre-trained on full-sequence (single-shot) offline trajectories with a standard cross-entropy loss to establish a robust spatiotemporal feature extractor. This backbone is subsequently fine-tuned using the online Fishbone strategy on continuously running circuits, where the quantum trajectories are generated interactively with the decoder's replenishment actions.

Both training stages utilize the AdamW optimizer ($\beta_1=0.9, \beta_2=0.98$) with a cosine learning rate decay and gradient clipping to ensure stable convergence. 

During inference, the replenishment action is deterministic, triggered when the predicted loss probability exceeds a threshold $\tau_d$: $u_{j,r+1} = \mathbb{I}\!\left[p_{j,r}^{\mathrm{loss}} \geq \tau_d\right]$. Instead of a naive $0.5$ threshold, $\tau_d$ is calibrated to explicitly navigate the trade-off between aggressive replenishment (high false-positive resets) and conservative waiting (prolonged defects). We optimized $\tau_d$ by maximizing the logical accuracy at round 20 on a separate validation set, yielding optimal thresholds of $\tau_3=0.86$, $\tau_5=0.78$, and $\tau_7=0.56$ for code distances 3, 5, and 7, respectively.

\subsection{Statistical Evaluation Metrics}

During evaluation, the reported logical accuracy is computed conditionally on surviving logical representatives. Let $m_{ikr} \in \{0,1\}$ indicate whether logical representative $k$ survives (has no lost qubits) in episode $i$ at round $r$, and let $\hat y_{ikr}$ and $y_{ikr}$ be its prediction and ground-truth label, respectively. The pooled conditional accuracy is defined as:
\begin{equation}
    A_r = \frac{\sum_{i,k} m_{ikr} \mathbb{I}[\hat y_{ikr} = y_{ikr}]}{\sum_{i,k} m_{ikr}}.
\end{equation}
The standard error (SE) is estimated with episodes as independent sampling units:
\begin{equation}
    \widehat{\mathrm{SE}}(A_r) = \frac{s_i(C_{ir} - A_r M_{ir})}{\sqrt{n}\,\overline{M}_r},
\end{equation}
where $M_{ir}$ and $C_{ir}$ are the counts of valid and correctly predicted representatives in episode $i$, $s_i$ is the sample standard deviation, and $n$ is the number of episodes.

To quantify the degradation of the hardware, we also define the physical-survival fraction $S_r$, representing the average proportion of physical qubits (both data and ancilla) remaining in the system prior to the next replenishment cycle:
\begin{equation}
    S_r = \frac{1}{n}\sum_i \frac{N^{\mathrm{present}}_{\mathrm{data},ir} + N^{\mathrm{present}}_{\mathrm{ancilla},ir}}{2d^2-1}.
\end{equation}
This macroscopic metric provides a direct visualization of how well the AI policy bounds the accumulation of physical defects.

\section{Decoding Graph Construction of Loss Events}
The delayed-erasure MWPM decoder used in this work was originally developed in~\cite{baranes2026leveraging}. Since the effects of ancilla loss are comparatively straightforward to analyze, we use data-qubit loss as an illustrative example to describe the construction of the decoding graph. Loss of a single data qubit can induce flips on neighboring ancilla checks due to non-commuting measurements. Under the measurement ordering adopted here, this non-commutativity gives rise to two classes of correlated syndrome-flip errors:
\begin{itemize}
  \item \textit{In-round}: The upper-right and lower-left checks of the same round flip simultaneously with probability $\frac{1}{2}$.
  \item \textit{Cross-round}: The upper-left check in round $r$ and the lower-right check in round $r+1$ flip simultaneously with probability $\frac{1}{2}$.
\end{itemize}

These rules specify the flip events induced by data-qubit loss in the bulk, which correspond to a two-endpoint edge in the decoding graph. In particular, the decoding graph features three types of boundaries: (i) spatial boundaries, (ii) the temporal boundary associated with the tail-round measurements, and (iii) the temporal boundary at which the loss event occurs. At these boundaries, the usual two-endpoint edge structure no longer applies and “dangling” endpoints can appear. For the first two boundary types, each dangling endpoint can flip independently at random; equivalently, the above two-check errors reduce to single-check flip errors. In contrast, at the loss-time boundary the dangling endpoint arising from cross-round correlations behaves differently: two additional edges are introduced between this endpoint and the two endpoints of the first cross-round edge in the bulk, enforcing the dangling endpoint flips simultaneously with exactly one of the latter endpoints.

Using the rules above, we construct the decoding graph for a single data-qubit loss event by adding the corresponding edges to the decoding graph induced by Pauli errors. In the experiment, data-qubit loss information must be delayed (or even unknowable); therefore, the full MWPM decoding graph adopts the same approximation as in~\cite{baranes2026leveraging}: we take a weighted average over all possible loss times and neglect interactions between multiple lost data qubits.

\end{document}